\documentclass[12pt,a4paper]{article}
\usepackage[english]{babel}
\usepackage{graphicx}

\newcommand{\alt}{\mathbin{\lower 3pt\hbox
   {$\rlap{\raise 5pt\hbox{$\char'074$}}\mathchar"7218$}}}
\newcommand{\agt}{\mathbin{\lower 3pt\hbox
   {$\rlap{\raise 5pt\hbox{$\char'076$}}\mathchar"7218$}}}

\begin{document}
\setcounter{footnote}{0}
\setcounter{equation}{0}
\setcounter{figure}{0}
\setcounter{table}{0}
\vspace*{5mm}

\begin{center}
{\large\bf Strong coupling asymptotics \\
of the $\beta$-function in $\varphi^4$ theory and QED}

\vspace{4mm}
\vspace{4mm}
I. M. Suslov \\
Kapitza Institute for Physical Problems,
\\  Moscow, Russia \\
\vspace{1mm}
\end{center}

\begin{center}
\begin{minipage}{135mm}
{\bf Abstract } \\
The well-known algorithm for summing  divergent series is based
on the Borel transformation in  combination with the conformal
mapping. A modification of
this algorithm allows one to determine a strong coupling
asymptotics of the sum of the series through the values of the
expansion coefficients. An application of the algorithm to the
$\beta$-function of $\varphi^4$ theory leads to the asymptotics
$\beta(g)=\beta_\infty g^\alpha$ at $g\to\infty$, where
$\alpha\approx 1$ for space dimensions $d=2,3,4$.
The natural hypothesis arises, that the asymptotic
behavior is $\beta(g) \sim g$ for all $d$. Consideration of the
"toy" zero-dimensional model confirms the hypothesis and reveals
the origin of this result: it is related to a zero of a certain
functional integral. A generalization of this mechanism to the
arbitrary space dimensionality leads to the linear asymptotics
of $\beta(g)$ for all $d$. The same idea can be applied to QED
and gives the asymptotics $\beta(g)=g$, where $g$ is the running
fine structure constant. A relation to the "zero charge"
problem is discussed.
\end{minipage}
\end{center} 


\vspace{6mm}
\begin{center}
{\bf 1. Introduction}
\end{center}

It is commonly accepted that summing  divergent series
can give important and non-trivial information. It will be
demonstrated below that sometimes we can obtain even
more:  summation of the series allows to guess the exact result
and then this result can be proved.

Our main interest is a reconstruction of the Gell-Mann -- Low
function $\beta(g)$ for actual field theories from its divergent
perturbation expansion. We describe the summation procedure in
Sec.\,2 and illustrate it for the case of $\varphi^4$ theory in
Sec.\,3. The arising hypothesis on the linear asymptotics
$\beta(g)\propto g$ is tested in Sec.\,4 in the zero-dimensional
limit, while Sec.\,5 gives its justification for any
dimension $d\le 4$. The same idea is applied to QED in Sec.\,6.
Finally,  Sec.\,7 discusses some problems arising in
relation to the obtained results.

\vspace{6mm}
\begin{center}
{\bf 2. Summation procedure}
\end{center}

Let us consider the typical problem in field theory applications.
A certain quantity $W(g)$ is defined  by its formal
perturbation expansion
$$
W(g)= \sum\limits_{N=0}^{\infty} \,W_N (-g)^N
\eqno(1)
$$
in the powers of the coupling constant $g$. The coefficients
$W_N$ are given numerically and have the factorial asymptotics
at $N\to\infty$,
$$
W_N^{as}=c a^N \Gamma(N+b) \,,
\eqno(2)
$$
which is a typical result obtained by the Lipatov method
\cite{1}. One can see that the  convergence radius  for (1)
is zero.  The problem arises, can we make any sense of the
series (1) and  find $W(g)$ for arbitrary $g$.

The conventional treatment of the series (1) is based on the
Borel transformation
$$
W(g)=\int\limits_{0}^{\infty} dx e^{-x} x^{b_0-1} B(gx)\,,\qquad
\eqno(3)
$$
$$
B(z)=\sum\limits_{N=0}^{\infty} B_N (-z)^N\,,\qquad
B_N=\frac{W_N}{\Gamma(N+b_0)}\,,
\eqno(4)
$$
relating the function $W(g)$ with its Borel
transform $B(z)$, while $B(z)$ is
given by a series with a factorially improved convergence;
$b_0$ is an arbitrary parameter, which can be used for
optimization of the procedure. Under the proper conditions,
Eq.3 is an identity obtained by interchanging of summation
and integration and using a definition of the
gamma-function. In the general case, Eqs.3,4 give {\it a
definition} of the Borel sum for a series (1).
In what follows, we identify the function $W(g)$ with
the Borel sum of its perturbation  series. In the case of
$\varphi^4$ theory, it is possible to test a validity of such
identification in one and zero dimensions
\cite{4} and to prove the Borel summability  in two
and three dimensions \cite{500,501}.

It is easy to show that the Borel transform $B(z)$ has a
singularity at the point $z=-1/a$ (Fig.\,1,\,$a$)
determined by the parameter
$a$ in the Lipatov asymptotics (2). The series for $B(z)$
is convergent in the disk  $|z|<1/a$, while we should know it
on the positive semi-axis, in order to perform integration
in the Borel integral  (3); so we need an analytical
continuation of $B(z)$. Such analytical continuation is easy
if the coefficients $W_N$ are defined by a simple formula,
but it is a problem when they are given numerically.
\begin{figure}
\centerline{\includegraphics[width=6.1 in]{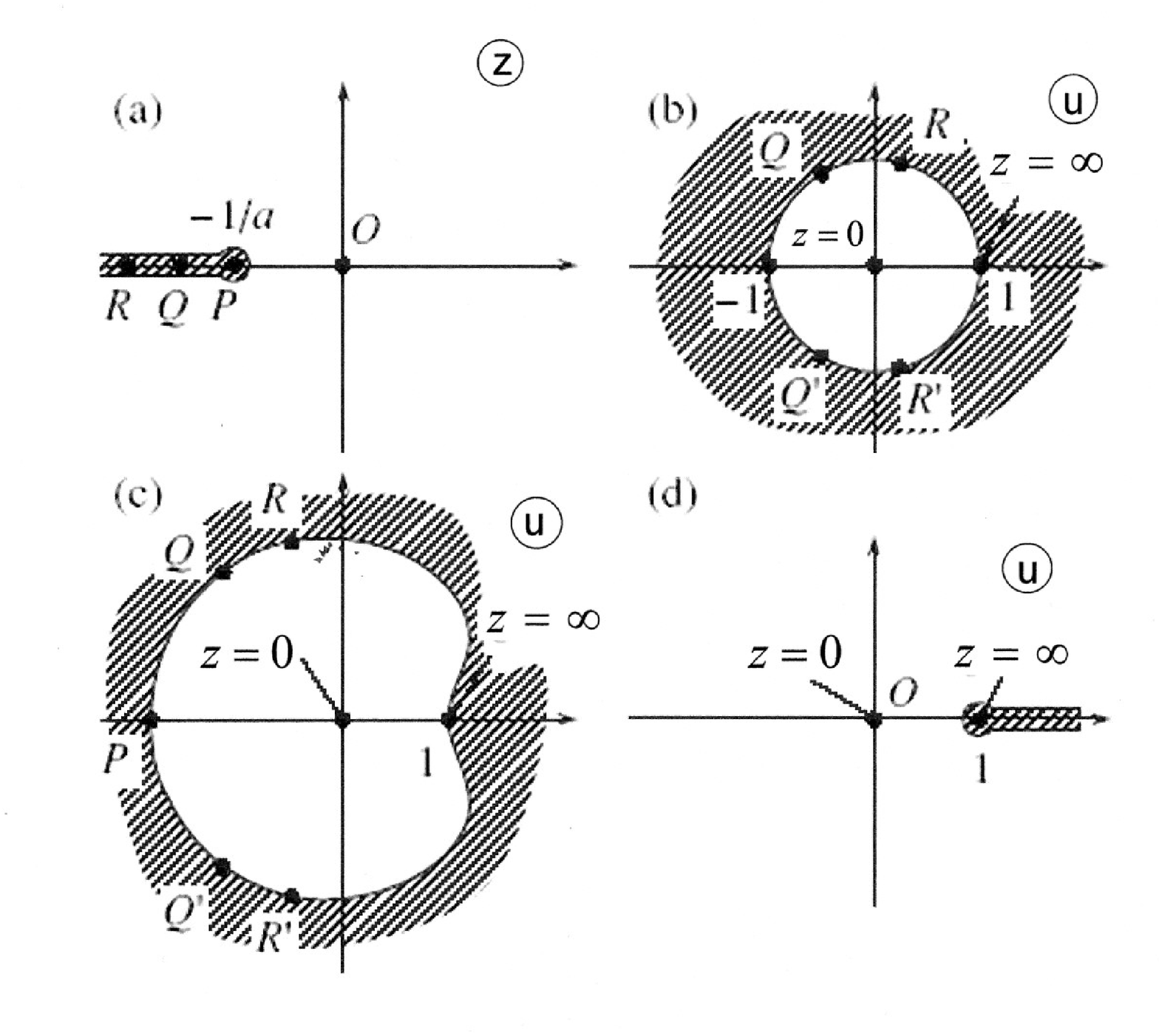}} \caption{
(a) The Borel transform $B(z)$ is analytical in the complex
plane with the cut $(-\infty, -1/a)$; (b) Its domain of
analyticity can be conformally mapped to a unit disk in the
$u$ plane; (c) If analytic continuation is restricted to the
positive semi-axis, then a conformal mapping can be made to any
domain, for which the point $u =1$ is the nearest to the origin of
all boundary points; (d) An extreme case of such domain is the
$u$ plane with the cut $(1,\infty)$. }
\label{fig1} \end{figure}

The elegant solution of this problem was given by Le Guillou and
Zinn-Justin in 1977 \cite{2}. It is based on the hypothesis that
in field theory applications all singularities of $B(z)$ lie on
the negative semi-axis.
This hypothesis can be proved in the case of $\varphi^4$ theory
\cite{3}.\,\footnote{\,A validity of this hypothesis
is frequently questioned in relation to possible
existence of the renormalon singularities  \cite{200}.
Such  singularities can be easily obtained by summing some special
sequences of  diagrams, but their existence was never
proved, if all diagrams are taken into account \cite{200a}.
The present results for the  asymptotics of the $\beta$-function
(Secs.\,5,\,6) are in agreement with a general criterion for
absence of renormalon singularities \cite{202} and a proof of
their absence  for $\varphi^4$ theory \cite{3} (see a detailed
discussion in \cite{201}).} If such analytical properties are
accepted, we can make a conformal transformation  $z=f(u)$,
mapping the complex plane with the cut (Fig.\,1,\,$a$) into the
unit disk $|u|<1$ (Fig.\,1,\,$b$). If we re-expand $B(z)$ in the
powers of $u$,
$$
B(z)=\sum\limits_{N=0}^{\infty} \left.B_N
(-z)^N\right|_{\displaystyle z=f(u)}\qquad
\longrightarrow\qquad
B(u)=\sum\limits_{N=0}^{\infty} U_N u^N \,\,,
\eqno(5)
$$
then such series will be convergent for any  $z$ except the cut
$(-\infty, -1)$. Indeed,
all singular points $P,\,Q,\,R,\ldots$ of  $B(z)$ lie on the
cut, and their images $P,\,Q,\,Q',\,R,\,R',\ldots$ in the $u$
plane appear on  the circle $|u|=1$. The
re-expanded series in (5) is convergent for $u$ lying
within the unit circle, but the interior of
the circle $|u|<1$ is in
one-to-one correspondence with
the analyticity domain in
the cutted $z$ plane (Fig.\,1,\,$a$).

Such conformal mapping is unique (apart from trivial
modifications), if we want to make an analytical continuation to
the whole domain of analyticity. In fact, such strong demand is
not necessary since we need $B(z)$ only at the positive semi-axis,
in order  to produce integration in (3). If we accept that the
image of  $z=0$ is  $u=0$ and the image of  $z=\infty$ is  $u=1$,
then we can make a conformal mapping to any domain, for which the
point $u=1$ is the nearest to the origin of all boundary points
(Fig.\,1,\,$c$). The series in $u$ converges for $|u|<1$, and in
particularly at the interval $0<u<1$, which is the image of the
positive semi-axis.


The advantage of such conformal mapping
consists in the possibility to
express the large $g$ asymptotics of $W(g)$ in
terms of the expansion coefficients $W_N$.
Indeed, the divergency of the series in $u$ is
determined by the nearest singular point  $u=1$, which is an
image of infinity: so the large $N$ behavior of the expansion
coefficients $U_N$ is related to the strong coupling
asymptotics of $W(g)$. In order to diminish influence of other
singular points $P,\,Q,\,Q',\ldots$, it desirable to remove
these points as far, as possible. Thereby, we come
to an extremal form of such conformal mapping, when it is made on
the whole complex plane with the cut  $(1,\infty)$
(Fig.\,1,\,$d$).  Mapping of the initial region (Fig.\,1,\,$a$)
to the region of Fig.\,1,\,$d$ is given by a simple rational
transformation
$$
z=\frac{u}{a(1-u)} \,,
\eqno(6)
$$
for which it is easy to find the relation of  $U_N$ and $B_N$,
$$
U_0=B_0\,,\qquad U_N=\sum\limits_{K=1}^{N} \frac{B_K}{a^K}
   (-1)^K C_{N-1}^{K-1}\qquad (N\ge 1)\,,
\eqno(7)
$$
where $C_N^K=N!/K!(N-K)!$ are the binomial coefficients.
If $W(g)$ has a power law asymptotics
$$
W(g)=W_\infty g^\alpha \, , \qquad g \to\infty  \,,
\eqno(8)
$$
then the large order behavior of $U_N$
$$
U_N=U_\infty N^{\alpha-1}\,,\qquad N\to \infty\,,
\eqno(9)
$$
$$
U_\infty=\frac{W_\infty}{a^\alpha \Gamma(\alpha) \Gamma(b_0+\alpha)}
\eqno(10)
$$
is determined by the parameters  $\alpha$ and $W_\infty$.
Consequently, we come to a very simple algorithm \cite{4}: the
coefficients  $W_N$  of the initial series  (1) define the
coefficients  $U_N$ of re-expanded series (5) according to
Eqs.\,4,\,7, while the behavior of $U_N$ at large $N$
(Eqs.\,9,\,10) is related to the strong coupling asymptotics
(8) of $W(g)$.

If information on the initial series (1) is sufficient for
establishing its strong coupling behavior (8), then
summation at arbitrary $g$ presents no problem. The
coefficients $U_N$ are calculated by Eq.7 for not very large
$N$, and then they are continued according to their asymptotics
(9).  Consequently, we know all coefficients of the convergent
series (5) and it can be summed with the required accuracy.

Few comments should be made to avoid a misunderstanding. The
conformal mapping corresponding to Fig.\,1,\,$b$ provides
(for fixed $z$)
the fastest convergence rate for the $u$ series  \cite{300}, and
is cited as "optimal" in the literature. It may look
preferable to use this algorithm and extract the asymptotics of
$W(g)$ from the summation results. In fact, all
investigators of the
strong coupling region \cite{203,204,205,4} independently came to
the same conclusion that the asymptotics of $W(g)$ should be
estimated {\it before} any summation.\footnote{\,For example, it
is clear from the described algorithm,
that one cannot find a correct asymptotics of $W(g)$, if
he does not  know a correct asymptotics  of  $U_N$.}
On the other hand,  the fastest convergence is
a distinctive excellence
only if $W_N$ are known exactly.
In the presence of round-off errors, the uncertainty in $U_N$
grows as $5.8^N$  for Fig.\,1,\,$b$ and as $2^N$  for
Fig.\,1,\,$d$ \cite{4}; more
than that, the latter (but not the former) algoritm is
stable in respect to smooth errors (like interpolation ones)
\cite{4}, and it
has a crucial significance for the following applications.

\vspace{6mm}
\begin{center}
{\bf 3. Application to $\varphi^4$ theory}
\end{center}

The described algorithm was successfully tested for a lot of
simple examples \cite{4}, and now we can apply it to a
reconstruction of the Gell-Mann -- Low function $\beta(g)$
of quantum field theories. This function enters the Gell-Mann
-- Low equation which describes the  behavior of the effective
charge $g$ as a function of the length  scale $L$:
$$
-\frac{dg}{d \ln L} =\beta(g)\,.
\eqno(11)
$$
The most interesting problem is an appearance
 of the $\beta$-function
in relativistic theories, like four-dimensional $\varphi^4$
theory or QED. In this case, the  expansion of
$\beta(g)$ begins with the positive quadratic term
and the  effective charge $g$ grows at small
distances\,\footnote{\,Equation (11) is valid
for $L\alt m^{-1}$, where $m$ is a mass of the particle;
in the region $L\agt m^{-1}$,  $g$ remains  constant and equal
to its observed value $g_{obs}$. }
(Fig.\,2):  it is interesting to find the law of this growth in
the strong coupling region.
\begin{figure}
\centerline{\includegraphics[width=5.1 in]{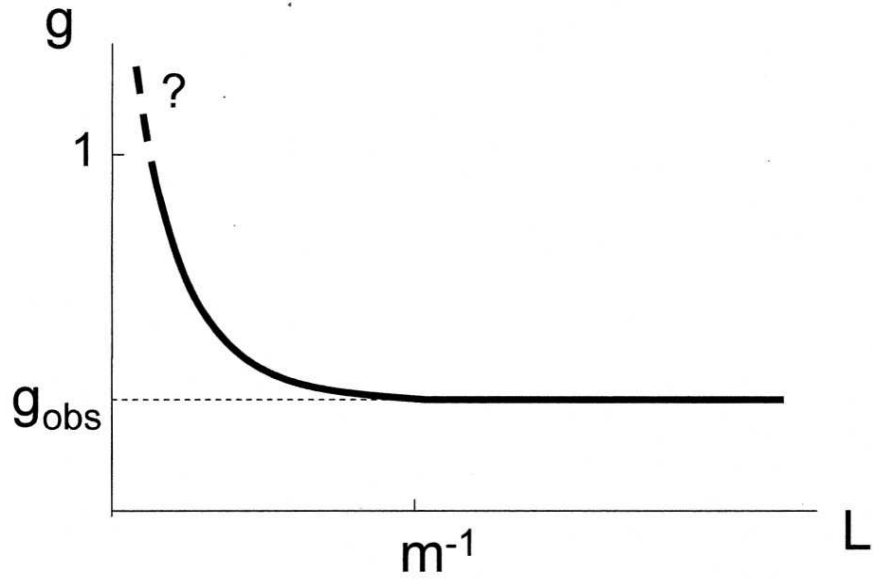}}
\caption{Effective coupling $g$ as a function of the length scale
$L$ in four-dimensional $\varphi^4$ theory and QED. }
\label{fig2}
\end{figure}

According to the classification by Bogolyubov
and Shirkov \cite{5}, there are three qualitatively different
possibilities (Fig.\,3):
\begin{figure}
\centerline{\includegraphics[width=6.5 in]{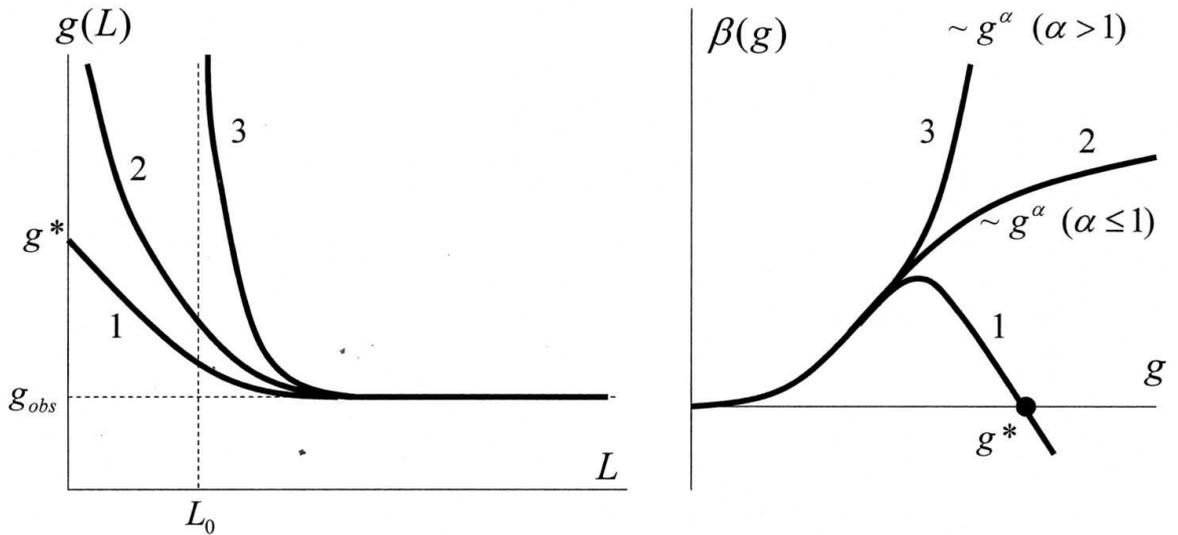}}
\caption{Three qualitatively different situations according to the
Bogolyubov and Shirkov classification. } \label{fig3}
\end{figure}
(1) if $\beta(g)$ has a  zero at some
point $g^*$, then the effective coupling  $g$ tends to
$ g^*$ at small $L$; (2) if $\beta(g)$ is
non-alternating and has the asymptotic behavior  $g^\alpha$
with $\alpha\le 1$, then $g(L)$ grows to
infinity; (3) if non-alternating $\beta(g)$ behaves at infinity
as $ g^\alpha$ with $\alpha>1$, then $g(L) $ is divergent  at
some finite $L_0$
and the  dependence  $g(L)$ is not defined at smaller distances:
the theory is internally inconsistent and a finite interaction
at large distances is impossible in the continual limit.
 To distinguish between these three possibilities, one needs
to know the $\beta$-function at
arbitrary $g$, and in particular its asymptotic behavior for
$g\to\infty$.

One can attempt to solve this problem by summation of the
perturbation series,
$$
\beta(g)= \beta_2 g^2+\beta_3 g^3+\ldots +\beta_L g^L
+\ldots+ c a^N \Gamma(N+b) g^N+ \ldots\,,
\eqno(12)
$$
having in mind that several first coefficients (till $\beta_L$)
are known from diagrammatic calculations and their large order
behavior is given by the Lipatov method. The intermediate
coefficients can be found by interpolation, the natural way for
which is as follows. It can be shown that corrections to the
Lipatov asymptotics has a form of the regular expansion in $1/N$:
$$
\beta_N=c a^N \Gamma(N+b) \left\{
    1+\frac{ A_1}{N}+\frac{ A_2}{N^2}+\ldots+
    \frac{ A_K}{N^K} +\ldots  \right\}
\,.
\eqno(13)
$$
One can truncate this series and choose the retained coefficients
$A_K$ from correspondence with the first coefficients
$\beta_2,\ldots,\beta_L$; then the interpolation curve  goes
through the several known points and automatically reaches its
asymptotics.   To variate this procedure, one can re-expand the
series (13) in the inverse powers of $N-\tilde N$,
$$
\beta_N=c a^N \Gamma(N+b) \left\{
    1+\frac{\tilde A_1}{N-\tilde N}+\frac{\tilde A_2}{(N-\tilde N)^2}+\ldots+
    \frac{\tilde A_K}{(N-\tilde N)^K} +\ldots  \right\}
\,,
\eqno(14)
$$
and obtain  a set of interpolations, determined by the arbitrary
parameter $\tilde N$.

In the case of four-dimensional $\varphi^4$ theory, a
realization of this program \cite{4} gives the
non-alternating $\beta$-function  (Fig.\,4,\,$a$), with
the results for the exponent $\alpha$ shown in
Fig.\,4,\,$b$.
\begin{figure}
\centerline{\includegraphics[width=6.1 in]{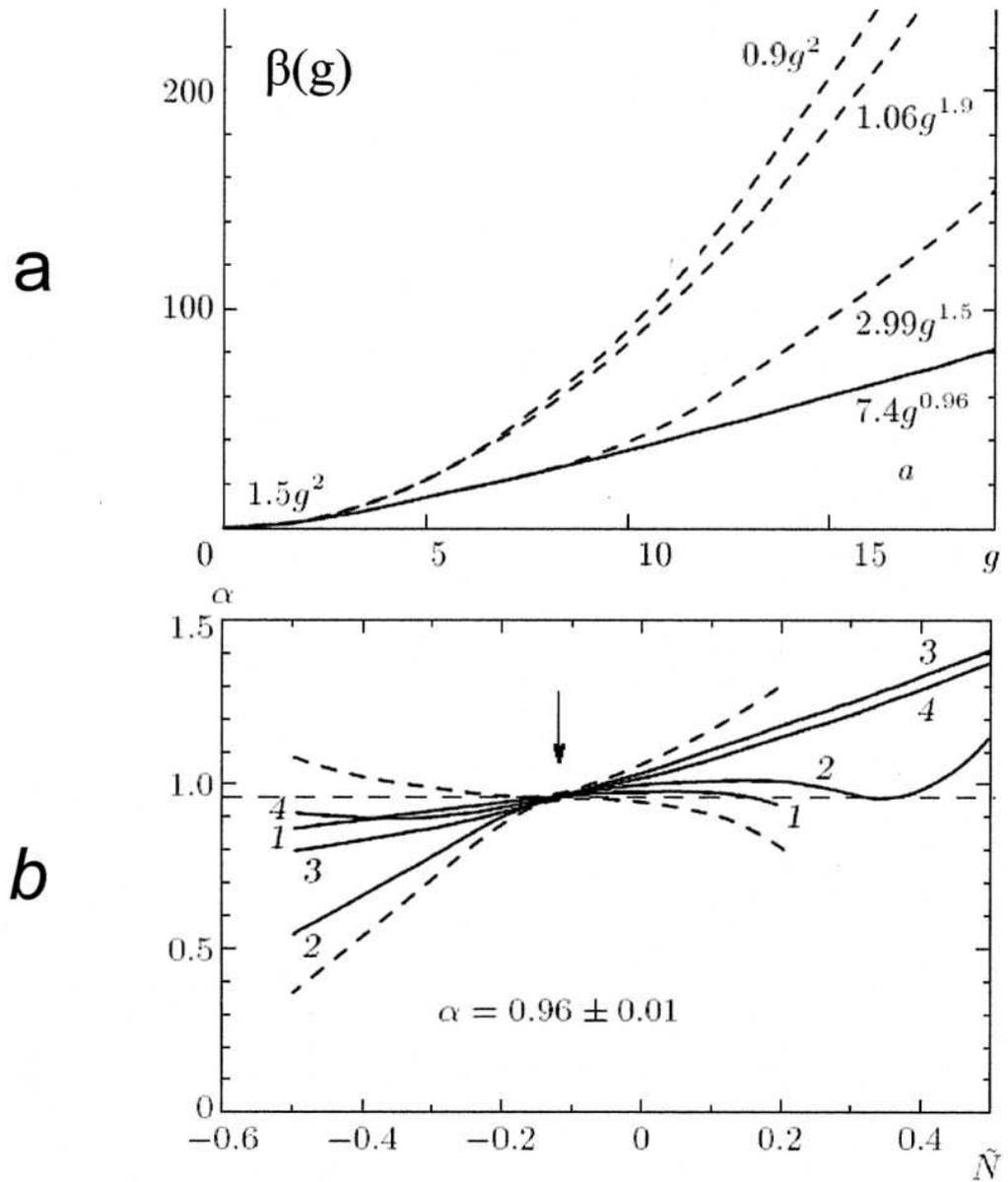}}
\caption{(a)
General appearance of the $\beta$-function in four-dimensional
$\varphi^4$ theory according to \cite{4} (solid curve), and
results obtained by other authors (upper, middle, and lower
dashed curves correspond to  \cite{203,204,205} respectively).
(b) Different estimations of the exponent $\alpha$ according to
\cite{4}. }
\label{fig4}
\end{figure}
The exponent $\alpha$ is
practically independent on $\tilde N$, and only its uncertainty
depends on this parameter. If we take the result with the minimal
uncertainty, we have a value $\alpha=0.96\pm 0.01$, surprisingly
close to unity.\footnote{\,Estimation of errors was made in
a framework of a certain procedure worked out in \cite{4}.
Subsequent applications have shown that such estimation is not
very reliable. }

Something close to unity is obtained also in two and three
dimensions \cite{6,7} (Fig.\,5).
\begin{figure}
\centerline{\includegraphics[width=5.1 in]{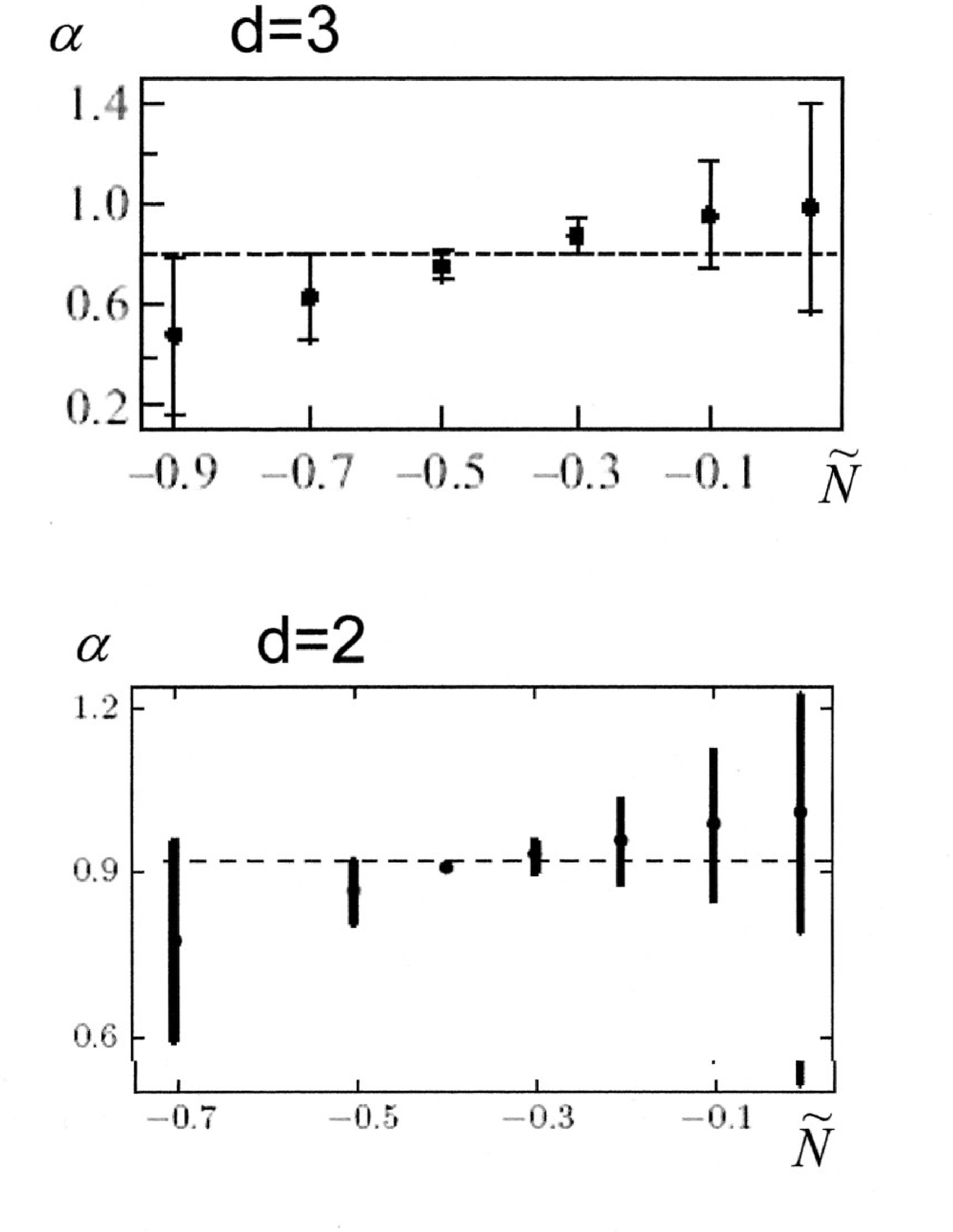}}
\caption{Estimations of the exponent $\alpha$
for $\varphi^4$ theory  in two and three
dimensions \cite{6,7}. } \label{fig5}
\end{figure}
The natural
hypothesis arises, that $\beta(g)$ has the linear asymptotics
$$
\beta(g)\sim g\,\,,\qquad g\to\infty
\eqno(15)
$$
for arbitrary space dimension  $d$. If this hypothesis is
correct, then there is a natural strategy for its
justification:

(i) to test it in a simple case $d=0$;

(ii) to find out the mechanism leading to this asymptotics;

(iii) to generalize this mechanism for arbitrary $d$.

\noindent
Surprisingly, this program can be realized and Eq.15 is
our main result. Since summation of the series gives
non-alternating $\beta(g)$  (Fig.\,4,\,$a$), we may conclude
that the second possibility of the Bogolyubov and Shirkov
classification is realized.

\vspace{6mm}
\begin{center}
{\bf 4. "Naive" zero-dimensional limit }
\end{center}

Consider the $O(n)$-symmetric  $\varphi^4$ theory with an
action
$$
S\{\varphi\} =\int \,d^dx \left\{
{\textstyle\frac{1}{2}} \sum_{\alpha=1}^n (\nabla
\varphi_\alpha)^2 + {\textstyle\frac{1}{2}} m_0^2
\sum_{\alpha=1}^n
\varphi_\alpha^{\,2} + {\textstyle\frac{1}{8}} u
\left(\sum_{\alpha=1}^n\varphi_\alpha^{\,2}\right)^2 \right\}\,,
$$
$$
u_0=g_0\Lambda^{\epsilon}\,, \qquad \epsilon=4-d
\eqno(16)
$$
in $d$--dimensional space; here $m_0$ is a bare mass, $\Lambda$
is a momentum cut-off, $g_0$ is a dimensionless bare charge.
It will be essential for us, that the $\beta$-function can be
expressed in terms of the functional integrals. The general
functional integral of $\varphi^4$ theory
$$
Z^{(M)}_{\alpha_1\ldots \alpha_M}(x_1,\ldots, x_M)=
\int D\varphi\,
\varphi_{\alpha_1} (x_1) \varphi_{\alpha_2} (x_2) \ldots
\varphi_{\alpha_M} (x_M) \exp\left(-S\{\varphi \} \right) \,
\eqno(17)
$$
contains $M$ factors of $\varphi$ in the pre-exponential; this
fact is indicated by the subscript $M$.

We can take a zero-dimensional limit, considering the system
restricted spatially in all directions. If its size is
sufficiently small, we can neglect the spatial dependence of
$\varphi(x)$ and omit the terms with gradients in Eq.17;
interpreting the functional integral as a multi-dimensional
integral on a lattice, we can take the system sufficiently small,
so that it contains only one lattice site. Consequently, the
functional integrals transfer to the ordinary integrals:
$$
Z^{(M)}_{\alpha_1\ldots \alpha_M} =
\int d^n\varphi\,
\varphi_{\alpha_1} \ldots
\varphi_{\alpha_M}
\exp\left(-{\textstyle\frac{1}{2}} m_0^2 \varphi^2-
{\textstyle\frac{1}{8}} u \varphi^{4}   \right) \,.
\eqno(18)
$$
This is the usual understanding of zero-dimensional theory.
Such model allows to calculate any quantities with zero external
momenta. If external momenta are not zero, the model is not
complete: it does not allow to calculate the momentum
dependence. To have a closed model, let us accept that there
is no momentum dependence at
all\,\footnote{\,This point is essential for
evaluation of the
$Z$-factor, which is defined in terms of the pair correlator
$G(x-x')=\langle \varphi(x)\varphi(x')\rangle$ in the momentum
representation as
$$
G(p)=\frac{1}{p^2+m_0^2+\Sigma(p,m_0)}
\equiv\frac{Z}{p^2+m^2+O(p^4)} \,,
$$
and is determined by the momentum dependence of self-energy.
In the described "naive" theory we accept  $Z=1$, since
the momentum dependence is absent.}.
 This "naive" model is
internally consistent but does not correspond to the
true zero-dimensional limit of $\varphi^4$ theory. The latter
fact is not  essential for us, since this model is used only for
illustration and the proper consideration of the general
$d$-dimensional case will be given in the next section.

Expressing the $\beta$-function in terms of functional
integrals, we obtain it in a form of the parametric
representation
$$
g=1- \frac{n}{n+2} \frac{K_4 K_0}{K_2^2}
\eqno(19)
$$
$$
\beta=-\frac{2n}{n+2} \frac{K_4 K_0}{K_2^2}
\left[ 2+\frac{ \frac{K_6 K_0}{K_4 K_2}-1 }
{ 1-\frac{K_4 K_0}{K_2^2}} \right] \,.
\eqno(20)
$$
The right hand sides of these formulas contain the
integrals
$$
K_M(t) = \int_0^\infty \varphi^{M+n-1} d\varphi\,
\exp\left(-t \varphi^2- \varphi^{4}   \right) \,,
\qquad t=\left(\frac{2}{u} \right)^{1/2} \,m_0^2 \,
\eqno(21)
$$
obtained from (18) by  simple transformations.
According to (19,\,20), the quantities $g$ and $\beta$ are
functions of the single parameter $t$; excluding $t$ we
obtain the dependence $\beta(g)$.

Investigation of  (19,\,20) for real $t$ shows
that  $g$ and $\beta$ as functions of $t$ have a
behavior shown in Fig.\,6,\,$a$; combination of these
results shows that  $\beta(g)$ behaves as in Fig.\,6,\,$b$.
\begin{figure}
\centerline{\includegraphics[width=5.5 in]{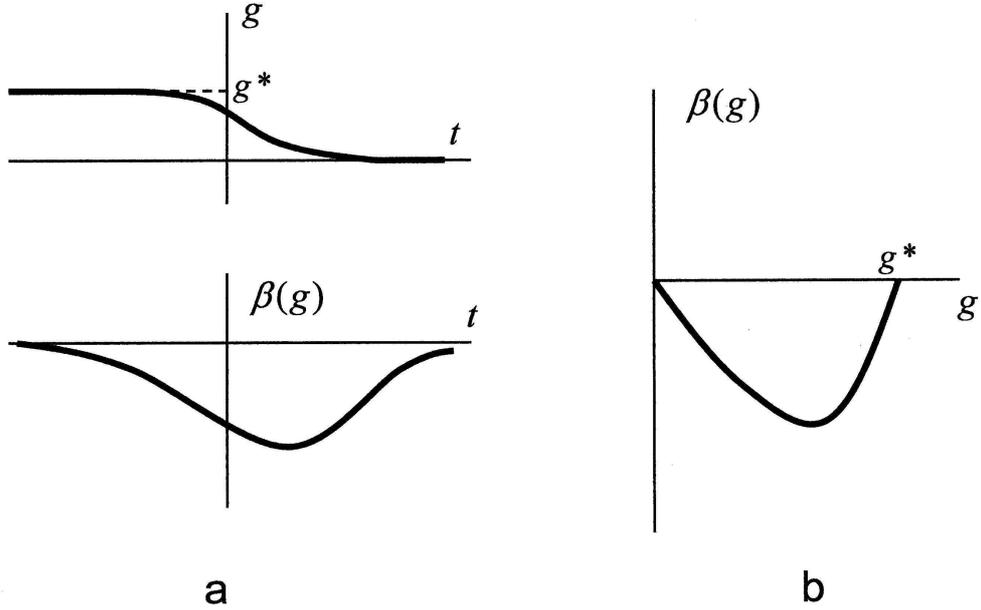}} \caption{a
---  Dependence of  $g$ and $\beta(g)$ on the parameter $t$. b ---
Resulting appearance of $\beta(g)$. } \label{fig6}
\end{figure}
We see that  variation of the parameter $t$ along the real axis
determines $\beta(g)$ in the finite interval $0\le g\le g^*$,
where $g^*$ is a fixed point\,\footnote{\,Existence
of the fixed point $g^*$
(obtained previously in \cite{301}) does
not mean the existence of a phase transition, which is absent
for  $d<2$ due to a finiteness of $m^2$. }
$$
g^* = \frac{2}{n+2}\,\,.
\eqno(22)
$$
To advance into the large $g$
region, we should consider the  complex values of $t$.

It appears, that in the complex $t$ plane  we should be
interested in zeroes of the integrals $K_M(t)$. The origin
of these zeroes is very simple. There are two saddle points
in the integral $K_M(t)$, the trivial and
nontrivial,
$$
\varphi_{c1}=0\,,\qquad \varphi_{c2}=\sqrt{-t/2}\,,
\eqno(23)
$$
and  $K_M(t)$ can be presented as a sum of two saddle point
contributions:
$$
K_M(t)= A_1 {\rm e}^{i\psi_1} + A_2 {\rm e}^{i\psi_2} \,.
 \eqno(24)
$$
If these two contributions compensate each other, then
the integral can turn to zero. Such compensation can be obtained
by adjustment of the complex parameter $t$, and in fact there are
infinite number of zeroes lying close to  lines
$\arg t=\pm 3\pi/4$ and accumulating at infinity (Fig.\,7).
\begin{figure}
\centerline{\includegraphics[width=4.5 in]{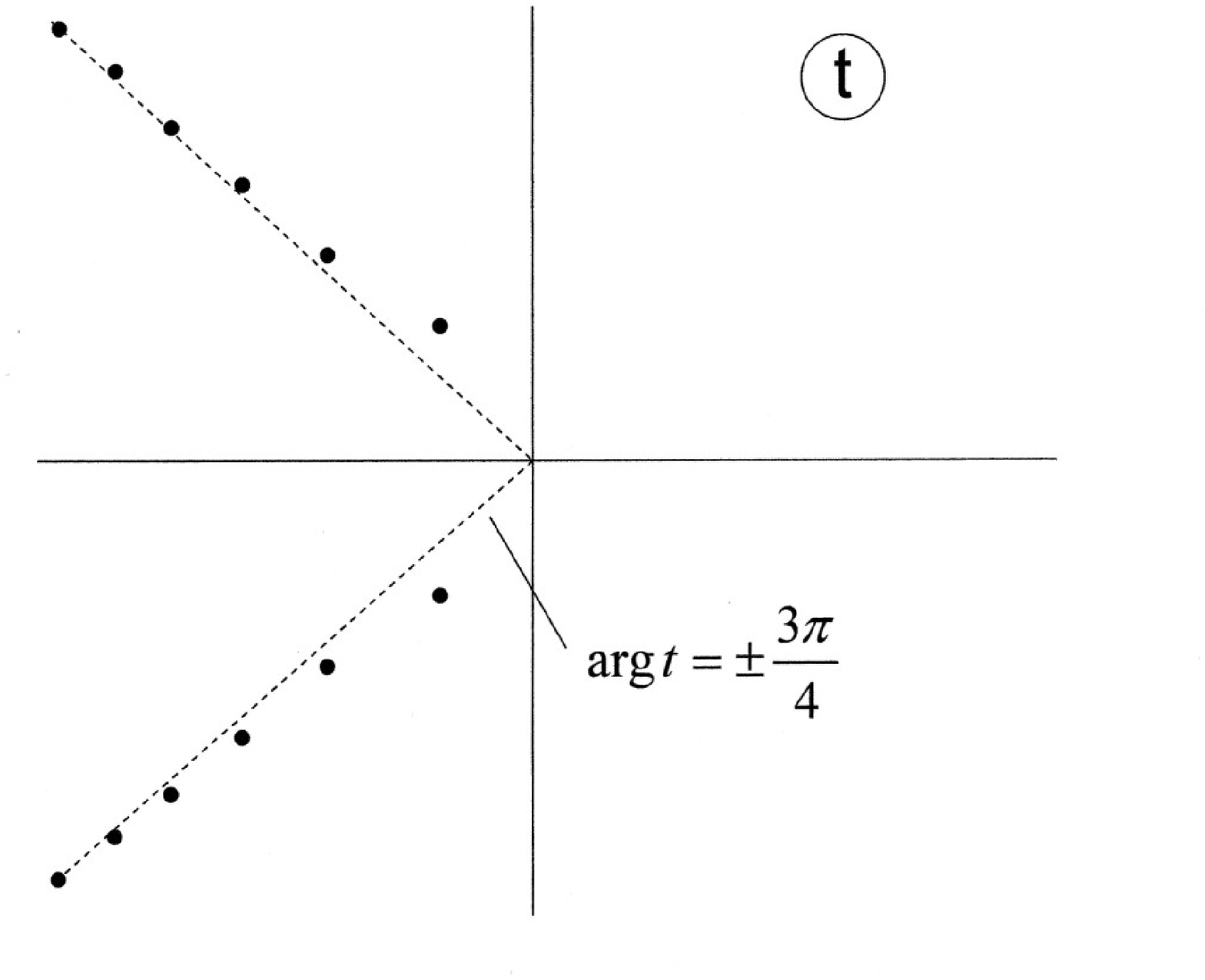}}
\caption{Zeroes of the integrals $K_M(t)$ in the complex $t$
plane. } \label{fig7}
\end{figure}
The above  saddle-point  considerations can be rigorously
justified for zeroes lying in the large $|t|$ region.
In fact, it is only essential for us that (i) zeroes of
$K_M(t)$ exist in principle, and (ii) zeroes of different
integrals  lie in different points.

Now return to the parametric representation (19,\,20). It
appears, that large values of $g$ can be achieved only
near the root of the integral  $K_2$. If $K_2$ tends to zero,
then  (19,\,20) are simplified,
$$
g\approx - \frac{n}{n+2} \frac{K_4 K_0}{K_2^2} \,,\qquad
\beta(g)\approx - \frac{4n}{n+2} \frac{K_4 K_0}{K_2^2}\,,
\eqno(25)
$$
and the parametric representation is resolved in the form
$$
\beta(g)= 4 g\,, \qquad g\to\infty \,.
\eqno(26)
$$
We see that, indeed, the asymptotic behavior of $\beta(g)$
appears to be linear.

\vspace{4mm}
\begin{center}
{\bf 5. General $d$-dimensional case} \\
\end{center}
\vspace{3mm}

The same ideas can be applied to the general $d$-dimensional
case. First of all, the actual functional integrals can turn to
zero by the same reason. Indeed,  the complex values of $t$
with large  $|t|$ correspond to complex  $g_0$ with small
$|g_0|$ (see Eq.21), and we come to a miraculous conclusion:
large values of the renormalized charge $g$ corresponds not to
large values of the bare charge $g_0$ (as naturally to
think\,\footnote{\,It is commonly accepted that the bare charge
$g_0$ is the same quantity as the renormalized charge $g$ at the
length scale $\Lambda^{-1}$.  In fact, these two quantities
coincide only on the two-loop level \cite{9} and this relation is
valid only in the weak coupling region.  }), but to its complex
values; more than that, it is sufficient to consider the region
$|g_0|\ll 1$, where the saddle-point approximation is applicable.
As a result,  the zeroes of the functional integrals can be
obtained by the compensation of the saddle-point contributions of
trivial vacuum and of the instanton configuration with the
minimal action; contributions of higher instantons are
inessential  for $|g_0|\ll 1$.

\vspace{3mm}

Now we need a representation of the $\beta$-function
in terms of functional integrals. The Fourier transform of
(18) will be denoted as $K_M$ after extraction of the
$\delta$-function of the momentum conservation and
a factor $I_{\alpha_1 \ldots \alpha_M}$ depending
on tensor indices:
$$
Z^{(M)}_{\alpha_1\ldots \alpha_M}(p_i)=K_M(p_i)\,
 I_{\alpha_1 \ldots \alpha_M}\,
 {\cal N}\delta_{p_1+\ldots+p_M}
 \eqno(27)
$$
where ${\cal N}$ is the number of sites on the lattice,
and $I_{\alpha_1 \ldots \alpha_M}$ is a sum of terms like
$\delta_{\alpha_1\alpha_2} \delta_{\alpha_3\alpha_4} \ldots$
with all possible pairings. In general, integrals $K_M(p_i)$
are taken at zero momenta, and only the integral
$K_2$ should be known for small momentum
$$
K_2(p)=K_2-\tilde K_2 p^2+\ldots
\eqno(28)
$$
Expressing the $\beta$-function in terms of functional
integrals\,\footnote{\,Definition of the $\beta$-function
depends on
the specific renormalization scheme. We accept renormalization
conditions at zero momenta (see Sec.VI.\,A in \cite{8}).},
we have a  parametric representation (see \cite{13} for
details):
$$
g=-\left(\frac{K_2}{\tilde K_2} \right)^{d/2}
\frac{K_4 K_0}{K_2^2}     \,,
\eqno(29)
$$
$$
\beta=\left(\frac{K_2}{\tilde K_2} \right)^{d/2}
\left\{ -d \frac{K_4 K_0}{K_2^2}
+2 \frac{(K'_4 K_0+K_4 K'_0)K_2 -2K_4 K_0 K'_2}{K_2^2}
\frac{\tilde K_2}{K_2\tilde K'_2-K'_2\tilde K_2 }
\right\}
\eqno(30)
$$
where the prime marks the derivatives over $m_0^2$.
If $g_0$ and $\Lambda$ are fixed, then the right hand sides of
these equations are functions of only $m_0$, while dependence on
the specific choice of $g_0$ and $\Lambda$ is absent due to
general theorems \cite{8}.

We see from Eq.29 that large values of  $g$ can be obtained
near the root of  either $K_2$, or $\tilde K_2$.
If $\tilde K_2\to 0$, equations  (29,\,30) are simplified,
so $g$ and $\beta$ are given by the same expression apart
from  a factor $d$,
$$
g=-\left(\frac{K_2}{\tilde K_2} \right)^{d/2}
\frac{K_4 K_0}{K_2^2} \,, \qquad
\beta=- d \left(\frac{K_2}{\tilde K_2} \right)^{d/2}
\frac{K_4 K_0}{K_2^2}
\,,\eqno(31)
$$
and the parametric representation is resolved as
$$
\beta(g)=d g\,,\qquad g\to \infty\,.
\eqno(32)
$$
For $K_2\to 0$, the limit $g\to\infty$ can be achieved
only for $d<4$ and we have analogously:
$$
\beta(g)=(d-4) g\,,\qquad g\to \infty\,.
\eqno(33)
$$
The results (32), (33)  correspond  to  different
branches of the analytical function  $\beta(g)$. It is easy to
understand that the physical branch is the first of them. Indeed,
it is well known from the phase transitions theory that
properties of $\varphi^4$ theory change smoothly
as a function of space dimension,
and results for $d=2,\,3$ can be obtained by
an analytic continuation from
$d=4-\epsilon$. According to all  available information,
the four-dimensional $\beta$-function is positive, and thus has
a positive asymptotics; by continuity, the positive asymptotics
is  expected for $d<4$.  The result (32) does obey these demands,
while
the branch (33) does not exist for $d=4$ at all.  Eq.\,32 agrees
with the approximate results discussed in Sec.\,3 and with the
exact asymptotic result $\beta(g)=2g$, obtained for
 the 2D Ising model \cite{20}
from the duality relation\,\footnote{\, Definition of the
$\beta$-function in \cite{20} differs by the sign from the
present paper.}.

\vspace{6mm}
\begin{center}
{\bf 6. Strong coupling asymptotics in QED}
\end{center}

The same ideas can be applied to QED.
Summation of perturbation series for QED \cite{10} gives
the non-alternating $\beta$-function (Fig.\,8)
\begin{figure}
\centerline{\includegraphics[width=5.1 in]{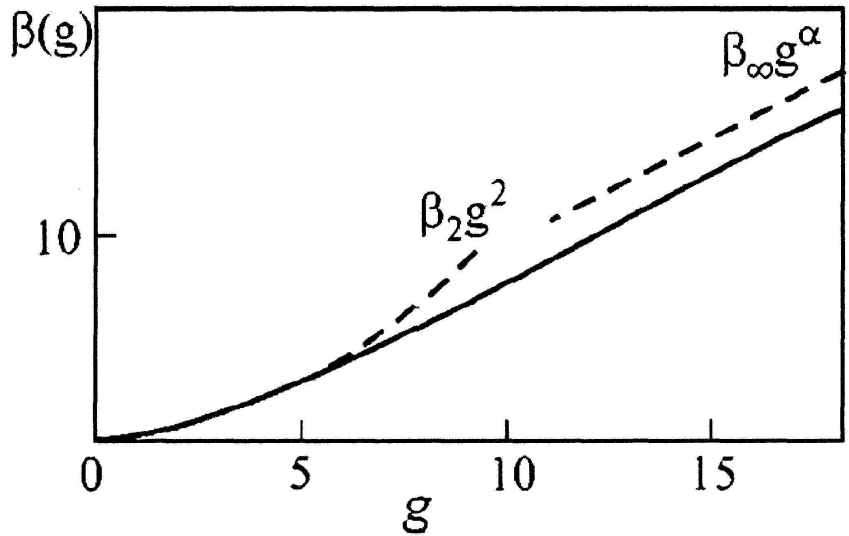}}
\caption{ General appearance of the $\beta$-function in QED
\cite{10}.} \label{fig8}
\end{figure}
with the asymptotics $\beta_\infty g^\alpha$, where (Fig.\,9)
\begin{figure}
\centerline{\includegraphics[width=7.0 in]{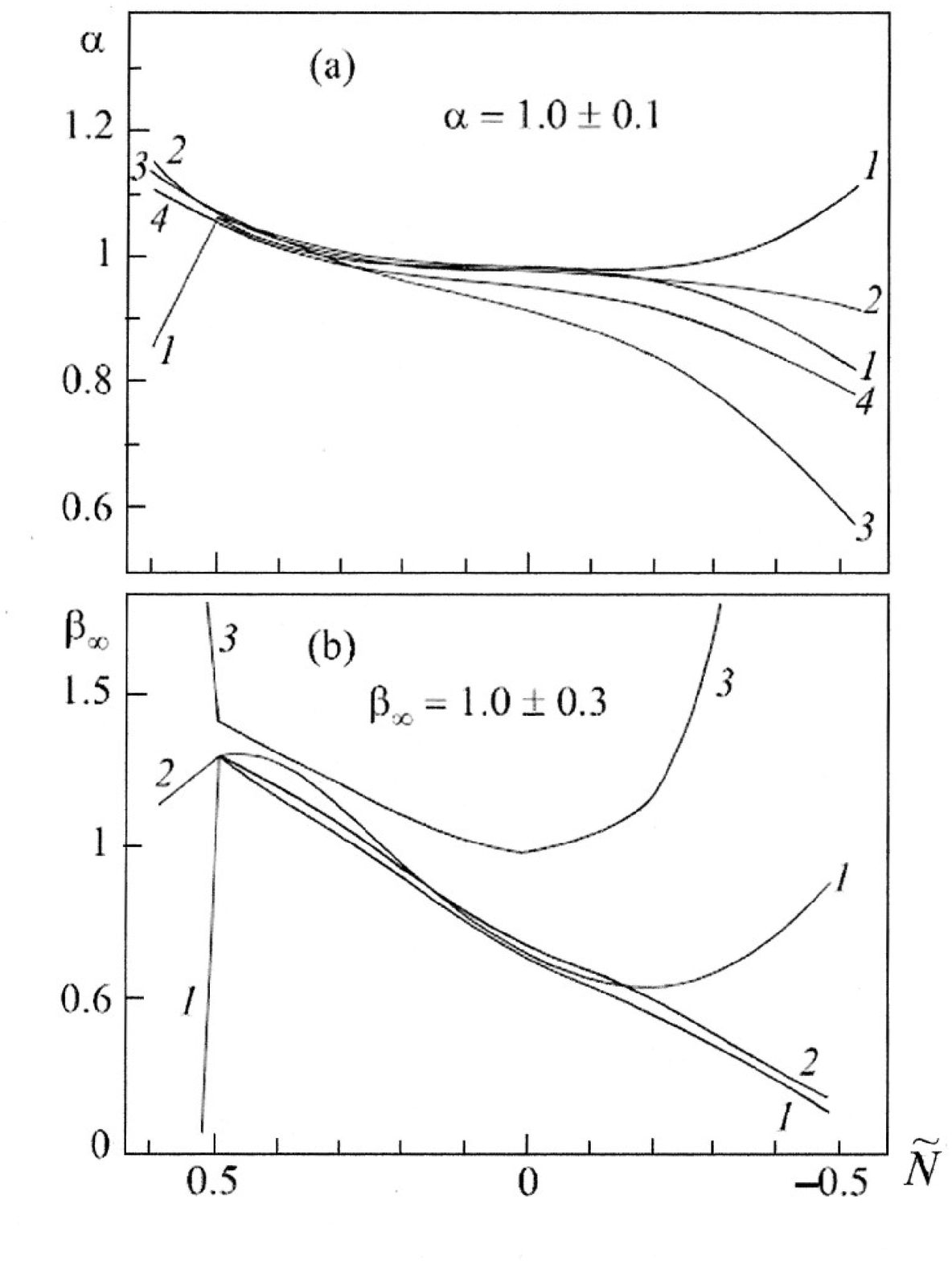}}
\caption{Different estimations of the parameters $\alpha$ and
$\beta_\infty$ for QED according to  \cite{10}. } \label{fig9}
\end{figure}
$$
\alpha=1.0\pm  0.1\,,\qquad \beta_\infty=1.0\pm 0.3 \,
\eqno(34)
$$
($g=e^2$ is the running fine structure constant).
 Within uncertainty, the obtained  $\beta$-function
satisfies inequality
 $$
 0\le \beta(g)< g \,,
 \eqno(35)
 $$
 established in \cite{11,12} from the spectral representations,
 while the asymptotics  (34) corresponds to the upper bound of
 (35).  Such coincidence does not look incident and indicates
 that the asymptotics $\beta(g)= g$ is an exact result.  We show
 below that it is so indeed.

The general functional integral of QED contains  $M$
photonic and $2N$  fermionic fields in the
pre-exponential,
  $$
  I_{M,2N}=\int DA D\bar\psi D\psi\, A_{\mu_1}(x_1)\ldots
  A_{\mu_M}(x_M)\,
          \psi(y_1)\bar\psi(z_1)\ldots \psi(y_N)\bar\psi(z_N)
\exp\left(-S\{A,\psi,\bar\psi\}  \right) \,,
  \eqno(36)
  $$
where $S\{A,\psi,\bar\psi\}$ is the Euclidean action,
$$
S\{A,\psi,\bar\psi\} =
\int d^4x \left[ \frac{1}{4}
               (\partial_\mu A_\nu- \partial_\nu A_\mu)^2
        +\bar\psi(i\!\!\not{\! \partial} -m_0 +
    e_0 \! \not  {\!\! A})\psi \right]\,,
  \eqno(37)
$$
while $e_0$ and $m_0$ are the bare charge and mass, and
the crossed symbols are convolutions of the corresponding
quantities with the Dirac matrices.  Fourier transforms of the
integrals $I_{M,N}$ with excluded $\delta$-functions of the
momentum conservation will be referred as $K_{MN}(q_i,p_i)$ after
extraction of the usual factors depending on tensor
indices\,\footnote{\,A specific form of these factors is
inessential, since the results are independent on the absolute
normalization of $e$ and $m$.}; $q_i$ and $p_i$ are momenta of
photons and electrons.

In general, these functional integrals are taken for zero
momenta, but two integrals $K_{02}(p)$ and $K_{20}(q)$ should be
estimated for small momenta: the first is linear in $p$,
and the second is quadratic in $q$,
 $$
K_{02}(p)=K_{02}+\tilde K_{02} \!\!\not{\!p}\,,\qquad
K_{20}(q)=K_{20}+\tilde K_{20} q^2\,,\qquad
\eqno(38)
$$
and in fact  the tilde denotes their momentum derivatives.

Expressing the $\beta$-function in terms of functional
integrals (see \cite{14} for details), we have a parametric
representation
$$
g= -\frac{K_{12}^2 K_{00}}{ \tilde K_{02}^2 \tilde K_{20}}\,,
\qquad
  \eqno(39)
$$
$$
\beta(g)= \frac{1}{2} \frac{K_{02} \tilde K_{02}}
{ K_{02}\tilde K'_{02} -K'_{02}\tilde K_{02} }
\frac{K_{12}^2 K_{00}}{ \tilde K_{02}^2 \tilde K_{20}}
\left\{ \frac{2\,K'_{12}}{ K_{12}} +\frac{K'_{00}}{ K_{00}}
-\frac{2\,\tilde K'_{02}}{\tilde K_{02}}
-\frac{\tilde K'_{20}}{\tilde K_{20}}
\right\}
  \eqno(40)
$$
where the prime denotes differentiation over $m_0$.
According to  Secs.4,\,5, the strong coupling regime for
renormalized interaction is related to a zero of a certain
functional integral. It is clear from (39) that the limit
$g\to\infty$ can be realized by two ways:  tending to zero either
$\tilde K_{02}$, or $\tilde K_{20}$.  For $\tilde K_{02}\to 0$,
equations (39,\,40) are simplified,
$$
g= -\frac{K_{12}^2 K_{00}}{ \tilde K_{02}^2 \tilde K_{20}}\,,
\qquad
\beta(g)= -\frac{K_{12}^2 K_{00}}{ \tilde K_{02}^2 \tilde
K_{20}}\,,
\eqno(41)
$$
and the parametric representation is resolved in the form
$$
\beta(g)= g\,,\qquad g\to \infty\,.
\eqno(42)
$$
For $\tilde K_{20}\to 0$, one has
$$
\beta(g)\propto g^2\,,\qquad g\to \infty\,.
\eqno(43)
$$
Consequently, there are two possibilities for the
asymptotics of  $\beta(g)$, either (42), or (43). The
second possibility is in conflict with inequality (35),
while the first possibility is in excellent agreement with
results (34) obtained by summation of perturbation series.
In our opinion, it is sufficient reason to consider
 Eq.42 as an exact result for the asymptotics of the
 $\beta$-function. It means that the fine structure
 constant in pure QED behaves as  $g\propto L^{-2}$ at small
 distances $L$.

\vspace{6mm}
\begin{center}
{\bf 7. Concluding remarks}
\end{center}

As should be clear from the preceeding discussion,
the conventional renormalization procedure defines  theory
only for $0\le g \le g_{max}$,  where $g_{max}$ is finite. For
values $g_{max}< g  <\infty$, the theory is defined by
an analytic continuation, and large values of
$g$ correspond to complex values of $g_0$.
Physically, the latter situation looks
inadmissible: the $S$-matrix can be expressed through the Dyson
$T$-exponential of the bare action, and Hermiticity of the bare
Hamiltonian looks crucial for unitarity of theory.

In fact, a situation is more complicated, as demonstrated by
Bogolyubov's axiomatical construction of the $S$-matrix \cite{5}:
according to it, the general form of the $S$-matrix is given by
the $T$-exponential of $iA$, where $A$ is a sum of (i) the bare
action, and (ii) a sequence of arbitrary "integration constants"
which are determined by quasi-local operators.  In the regularized
theory we can set the "integration constants" to be zero, and the
$S$-matrix is determined by the bare action. However, in the
course of renormalization these constants are taken non-zero, in
order to remove divergences.  These non-zero
"integration constants" can be absorbed by the action due to the
change of its parameters. As a result, for the true continual
theory the $S$-matrix  is determined by the renormalized action,
while the bare Hamiltonian  and the
Schr$\ddot o$dinger equation are
ill-defined. From this point of view there is no problem with the
complex bare parameters, since the renormalized Lagrangian is
Hermitian for real $g$.

Some problems remain for regularized theory, where the bare and
renormalized Lagrangians are equally admissible and a situation
looks controversial. The analogous situation
was discussed for the exactly solvable Lee model \cite{xx1}, which
also has the complex bare coupling for the sufficiently large
renormalized coupling. After the paper \cite{302} it was generally
accepted that the Lee model
is physically unsatisfactory due to existence of "ghost" states
(i.e. the states with a negative norm). Quite recently \cite{xx5}
it was found that this point of view is incorrect and the Lee
model is completely acceptable physical theory. It is a key idea
of \cite{xx5} that an analytical continuation
of the Hamiltonian parameters to the complex plane should be
assisted by a modification of the inner product for the
corresponding Hilbert space,
$$
(f,g)=\int f^*(x) g(x) dx \qquad  \longrightarrow\qquad
(f,g)_G= (f, \hat G g) \,,
$$
and with the proper choice of the operator $\hat G$ the bare
Hamiltonian is Hermitian in respect to the new inner product
$(f,g)_G$. As a result, all states of the Lee model have a
positive norm and evolution is unitary. The analogous procedure
should exist in the present case, in order to remove the
indicated controversy.  In fact, a definition of charge
is ambiguous due to
ambiguity of the renormalization scheme \cite{9}
(arising from arbitrariness of "integration constants"
in Bogolyubov's construction)
 and  complex-valuedness of
 $g_0$ has a relative sense (see Sec.5 of \cite{13}).

The result $\alpha=1$ corresponds to one of the really existing
branches of the $\beta$-function,  analytically
continued from the weak coupling region. Strictly speaking, we
did not prove that this branch is physical. This point, together
with complex-valuedness of $g_0$, casts certain doubt on the
physical relevance of this result. However, our approximate
summation results (Secs.3,\,6), the exact result for the Ising
model \cite{20} and inequality (35) for QED give the
essential evidence that the result $\alpha=1$ is physical.

\vspace{2mm}

\qquad\qquad\qquad\qquad\qquad\qquad----------------------
\vspace{2mm}

In conclusion, summation of perturbation series gives the
positive $\beta$-function in four-dimensional $\varphi^4$ theory
and QED, while its strong coupling asymptotics is shown to be
linear.  It means that the second possibility in the Bogolyubov
and Shirkov classification (Sec.3) is realized, and it is
possible to construct the continuous theory with finite
interaction at large distances.\,\footnote{\,Discussion of
a wide-spread opinion on "triviality" of $\varphi^4$ theory is
given in \cite{xx7}. }

\vspace{3mm}


\end{document}